# THE HAZARDS OF PROPPING UP: BUBBLES AND CHAOS

Philip Maymin, Polytechnic Institute of New York University

**ABSTRACT**

*In the current environment of financial distress, many governments are likely to soon become major holders of financial assets, but the policy debate focuses only on the likelihood and extent of short-term market stabilization. This paper shows that government intervention and propping up are likely to lead to long-term bubbles and even wildly chaotic behavior. The discontinuities occur when the committed capital reaches a critical amount that depends on just two parameters: the market impact of trading and the target exposure percentage.*

## INTRODUCTION

Globally, governments appear to be quickly becoming holders of enormous quantities of financial securities. The main goals are to calm the markets, increase liquidity, and raise prices. While the details are still unresolved, the vast majority of financial economists seem to support some sort of intervention. But the government's buying power, whether or not it leads to a short-term stabilization of prices, can in certain circumstances described by this paper lead to a long-term bubble, or even chaos.

Consider a financial institution that faces mark-to-market losses on its large position of illiquid financial instruments. It can "prop up," or make additional market transactions that are small relative to its overall portfolio but, because of the illiquidity, have a large effect on the market price and therefore on the value of its entire portfolio. In this way, the institution can appear to generate large profits from its existing position at a remarkably small cost.

This paper presents a model of such mark-to-market propping up for a stylized financial entity aiming to hold a position in thinly traded derivatives with an exposure in fixed proportion to its capital. The prices resulting from the entity's transactions show variety and complexity, ranging from traditional bubbles, to discontinuous gaps, to smooth hills, to infinite growth.

Funds often allocate a certain portion of their capital to a particular investment class. Whether it is a pension fund allocating half of its capital to bonds, a mutual fund allocating one-third of its capital to growth stocks, or a sovereign wealth fund allocating twenty percent of its capital to a particular currency, it is common to find fixed percentage targets for entire classes of securities, with the percentage changing relatively rarely.

Consider an example of a hedge fund that trades only volatility swaps, which pay off the difference between the realized volatility of an underlying asset and the initial strike, multiplied by the notional of the swap, called the vega notional. Suppose a $100 million hedge fund, which seeks to always be long 10 percent of its capital in vega notional, is indeed long $10 million worth of vega notional. It trades in fixed maturities of, let us suppose, five years. After one year has elapsed, the mark-to-market profit of the hedge fund is equal to the product of $10 million and the difference between the weighted average of the realized volatility for the first year and the implied volatility of the remaining four years. Because one-fifth of the position has expired, the remaining vega notional is now only $8 million. If the fund wants to maintain a 10% exposure to volatility, it needs to purchase some more volatility swaps.

But what if the market implied volatility is no longer cheap? Could it still make sense for the fund to purchase the swap? Yes, it could, if the fund is big enough to cause a market impact by its trading. Suppose that by purchasing $2 million of new five-year vega, the hedge fund causes all implied volatilities to rise by one volatility point above fair value. Then assuming the fund does no more trades, in five years it will lose $2 million on its last trade. In the meantime, however, it will have caused a mark-to-market profit of $8 million, because its old volatility swaps would have been marked to the higher value too. Over the next four years, it will lose all of that $8 million as well, since at the end, the realized volatility will be the fair value, and the marks at the end of the first year will be deemed, in hindsight, to be too high.

The rest of this paper contains a literature review, the formal model definitions and dynamics, and an exploration and discussion of the simulations, followed by a conclusion.

## LITERATURE REVIEW

The literature on bubbles is perhaps the oldest class of financial research. Aristotle wrote of Thales who created a bubble in olive presses by cornering the market and selling at its peak. Isaac Newton lost twenty thousand pounds in the South Sea bubble, famously complaining that he can "calculate the motions of heavenly bodies, but not the madness of people." These two early examples still represent the two primary modern approaches to understanding bubbles.

Explanations of Thales-like bubbles assume relatively rational investors. Blanchard and Watson (1982) show that bubbles can be consistent with rationality. Jarrow (1992) finds bubbles can form if certain large traders can manipulate markets. Abreu and Brunnermeier (2003) find bubbles can persist because of a synchronization problem between rational investors in timing their exits.

Explanations of South Sea-like bubbles assume the existence of noise traders, combined with limits on arbitrageurs. Shleifer and Summers (1990) show how even rational investors can predict bigger fools in the future and so push bubbles further in the hope that they will be able to get out before it bursts. Many psychological biases seem able to drive noise trader behavior. Scheinkman and Xiong (2003) show that overconfidence leads to bubbles. Greenwood and Nagel (2008) find that the recent tech bubble was driven by younger investors. Barberis, Shleifer, and Vishny (1998) show that naïve extrapolation and a conservatism bias lead to consistently predictable return patterns.

This paper does not assume a particular psychological bias for the financial entities. On the contrary, the entities act rationally given their constraints.

Indeed, regardless of whether market participants are rational or not, experimental markets reliably replicate bubbles in laboratory conditions. Porter and Smith (2003) review 72 such experiments and find that markets with a well-defined expected fundamental dividend value exhibit bubbles, which diminish with experience, and which seem to depend on investor uncertainty about the behavior of other traders rather than uncertainty about the fundamental value itself. Hirota and Sunder (2007) suggest alternatively that short horizons and computational difficulty are to blame.

It is possible to generate bubbles and crashes through explicit models of agent behavior. For example, Corcos, Eckmann, Malaspinas, Malevergne, and Sornette (2008) introduce a deterministic, infinite-agent model in which agents change from bullish to bearish depending on the mood of their friends. This model is able to generate a wide range of behavior from chaotic to super-exponentially growing bubbles followed by crashes to quasi-periodic behavior. Wolfram (2002) describes a very simple idealized model of a market based on all entities repeatedly changing their mind depending on the decisions made by their two nearest neighbors in the prior period. In a similar vein, Maymin (2008a) proposes the minimal model of complexity of financial security prices, requiring only a single investor trading a single asset, yet still generating bubbles, crashes, and complexity.

Relative to such agent-based models, the dynamics presented here do not depend on complexity arising from simple trading rules or changes in attitudes or beliefs, but rather focus on the results of the fixed portfolio allocation decisions of a single financial entity or group of financial entities.

Furthermore, agent-based rules are typically applied to asset markets such as stocks or bonds, eschewing the more complicated derivatives markets, whereas the model presented here applies most easily to derivatives with fixed maturity dates and random realizations rather than perpetuities like stocks or fixed coupons like bonds.

Indeed, research linking bubbles and derivatives is rare and often only an indirect link based on bubbles in the underlying asset. For example, Cox and Hobson (2005) show that option pricing in the presence of an underlying bubble violates put-call parity among other things. In contrast, this paper describes bubbles formed directly in derivative markets by the market participants.

Similarly, financial research on chaos such as Hsieh (1991) has focused on the dynamics of the underlying asset processes, not on the chaos potentially caused directly through trading in markets for the derivatives themselves.

## MODEL DEFINITIONS AND DYNAMICS

Let us define some terms.

**Definition 1:** A *standard maturity* derivative security is one that is typically traded in the market for a fixed maturity of a particular number of years. ■

The most obvious examples are credit default swaps, where new swaps tend to trade at five-year maturities and old swaps have maturities less than five-years. Volatility and variance swaps can typically be traded for any maturity but one-year and two-year maturities are common as well. Sometimes such swaps tend to expire on particular days such as to coincide with the expiration of a futures contract, so the maturity is a constant when rounded to the nearest year.

**Definition 2**: A *decomposable* derivative security is one whose payoff can be expressed as:

$$\sum_{t=0}^{T-\Delta t} \frac{N\Delta t}{T}\left(R_{t;t+\Delta t} - K\right)$$

where the derivative security has initial strike $K$ and notional $N$ and which pays off $N\left(R_{0;T} - K\right)$ at maturity $T$, with $R_{A;B}$ representing the realized value of the period from $A$ to $B$. ■

In other words, a decomposable derivative security is a sequence of forward-starting derivatives each with maturities $\Delta t$. A typical example of such a security is a variance swap.

**Definition 3**: A *linear market impact model* for a family of derivative securities with a flat term structure adjusts all future implied values by the same amount for a given amount of notional traded. In particular, for a variance swap, $\sigma' = \sigma + \lambda V$, where σ′ is the new implied level, $\sigma$ is the old implied level, $V$ is the amount of variance swap notional purchased (a negative amount if sold), and $\lambda$ is the impact. ■

In other words, and simplifying to the language of volatility swaps, if you buy \$10 million of vega, and if we assume as we will later that $\lambda = 0.05$, then you will have increased the remaining implied volatility by half of a volatility point. If it had been a flat 20% implied volatility term structure, it will now be a flat 20.50% implied volatility term structure.

**Definition 4**: A *constant exposure* fund is one that seeks to maintain an exposure to a standard maturity derivative equal to a fixed proportion of its capital at every period. ■

A fund that aims to be 60% invested in equities and 40% in bonds is a constant exposure fund if one regards stocks as derivatives whose standard maturity is infinity and bonds as a derivative whose standard maturity is 30 years. As the bond values accrete down, such a fund needs to purchase more new bonds to satisfy its exposure requirement.

A fund that aims to maintain a 10% exposure to volatility will rebalance its volatility exposure every period to ensure that any profit it has earned is used to support new and larger positions and any losses it has incurred will result in a sale of new positions to offset the exposure of its existing positions.

With these definitions, we can derive the evolution dynamics of a hedge fund's capital.

**Theorem 1:** A constant exposure fund (or group of funds acting as one) with a target constant proportional exposure of $\kappa$ to a decomposable derivative security that trades in the market with a standard maturity of $T$ years, having a linear market impact model that increases the implied mark of all derivative securities by $\lambda$ for each unit of exposure purchased, will, at time $t + \Delta t$, have capital $C_{t+\Delta t}$, average maturity $M_{t+\Delta t}$, and implied mark $\sigma_{t+\Delta t}$ given jointly by:

$$C_{t+\Delta t} = C_t + \frac{C_t \kappa \Delta t \left( R_{t;t+\Delta t} - \sigma_{t+\Delta t} + C_t \kappa \lambda \frac{(M_t - \Delta t)}{M_t} \right)}{M_t - C_t \kappa^2 \lambda (M_t - \Delta t)}$$

$$M_{t+\Delta t} = \frac{M_t T \Delta t \left[ \kappa \left( \sigma_t - R_{t;t+\Delta t} \right) - 1 \right] - (M_t - \Delta t)^2 \left( M_t - C_t \kappa^2 \lambda (M_t - \Delta t) \right)}{C_t \kappa^2 \lambda (M_t - \Delta t)^2 - M_t \left( M_t - \kappa \Delta t \left( \sigma_t - R_{t;t+\Delta t} \right) \right)}$$

$$\sigma_{t+\Delta t} = \frac{\sigma_t M_t + C_t \kappa \lambda \left( \Delta t - \kappa \sigma_t M_t + \kappa \Delta t R_{t;t+\Delta t} \right)}{C_t \kappa^2 \lambda (M_t - \Delta t) - M_t}$$

where $R_{t;t+\Delta t}$ is the realized portion of the derivative for the period from time $t$ to time $t + \Delta t$ and $C_0$ is the fund's initial starting capital.

**Proof of Theorem 1:** The proof follows from solving the following equations, which themselves follow directly from the assumptions and the definition of profit as the sum of the realized profit plus the implied profit:

| | |
|---|---|
| $V_t = \kappa C_t$ | "Vega" definition |
| $\tilde{V}_t = V_t \frac{M_t - \Delta t}{M_t}$ | "Aged vega" |
| $V_{t+\Delta t} = V_t + \kappa \Pi_{t;t+\Delta t}$ | New vega |
| $\sigma_{t+\Delta t} = \sigma_t + \lambda\left(V_{t+\Delta t} - \tilde{V}_t\right)$ | Market impact |
| $\Pi_{t;t+\Delta t} = \frac{V_t}{M_t}\left(R_{t;t+\Delta t} - \sigma_t\right)\Delta t + \tilde{V}_t(\sigma_{t+\Delta t} - \sigma_t)$ | Profit for this period |
| $M_{t+\Delta t} = \frac{\tilde{V}_t(M_t - \Delta t) + \left(V_{t+\Delta t} - \tilde{V}_t\right)T}{V_{t+\Delta t}}$ | Average maturity |
| $C_{t+\Delta t} = C_t + \Pi_{t;t+\Delta t}$ | Definition of profit |

Substituting the definitions of $\sigma_{t+\Delta t}$, $\widetilde{V}_t$, and $V_{t+\Delta t}$ from the earlier equations into the one for profit, we get:

$$\Pi_{t;t+\Delta t} = \frac{V_t}{M_t}\left(R_{t;t+\Delta t} - \sigma_t\right)\Delta t + \widetilde{V}_t(\sigma_{t+\Delta t} - \sigma_t)$$

$$= \frac{V_t}{M_t}\left(R_{t;t+\Delta t} - \sigma_t\right)\Delta t + \tilde{V}_t\lambda\left(V_{t+\Delta t} - \tilde{V}_t\right)$$

$$= \frac{V_t}{M_t}\left(R_{t;t+\Delta t} - \sigma_t\right)\Delta t + \tilde{V}_t\lambda\left(V_t + \kappa\Pi_{t;t+\Delta t} - \tilde{V}_t\right)$$

$$= \frac{V_t}{M_t}\left(R_{t;t+\Delta t} - \sigma_t\right)\Delta t + V_t\frac{M_t - \Delta t}{M_t}\lambda\left(V_t + \kappa\Pi_{t;t+\Delta t} - V_t\frac{M_t - \Delta t}{M_t}\right)$$

Collecting terms gives the solution for $\Pi_{t;t+\Delta t}$ from which all others flow:

$$\left(1 - V_t\frac{M_t - \Delta t}{M_t}\lambda\kappa\right)\Pi_{t;t+\Delta t} = \frac{V_t}{M_t}\left(R_{t;t+\Delta t} - \sigma_t\right)\Delta t + V_t^2\frac{M_t - \Delta t}{M_t}\lambda\frac{\Delta t}{M_t}$$

■

The "vega" definition in the proof represents the following. If the fund has \$100 million and is targeting a 10% exposure to volatility, then the amount of vega it is carrying is \$10 million. This follows from the definition of a constant exposure fund.

The "aged vega" is the amount of vega remaining after a period of time $\Delta t$ has elapsed. By decomposability, the progress of time has essentially expired the first $\Delta t$ of the total $M_t$ maturity of the decomposable derivative security. Hence the remaining exposure is just the fraction of the remaining maturity as applied to the initial exposure to the derivatives.

The profit for any period is composed of two parts: the realized profit on the portion of the decomposable derivative that has essentially matured, and the implied profit on the portion of the decomposable derivative that remains. The portion that has matured is $\frac{V_t}{M_t}\Delta t$ and the portion that remains is the "aged vega."

The average remaining maturity is a weighted average of the maturity that remained, weighted by the aged vega, and the standard maturity of the derivative, weighted by the new vega, or the additional amount of vega required to be purchased to maintain the constant-exposure assumption. Since the new capital is simply the sum of the old capital and the profits, the new vega must be the same constant proportion $\kappa$ of the new capital. Since the old vega was that same proportion of the old capital, the new vega can be expressed as the sum of the old vega plus a proportion $\kappa$ of the profits.

Note the discontinuities that can result when the coefficient of the left hand side of the equation for $\Pi_{t;t+\Delta t}$ is near zero:

$$V_t = \frac{M_t}{M_t - \Delta t}\frac{1}{\lambda\kappa}C_t = \frac{M_t}{M_t - \Delta t}\frac{1}{\lambda\kappa^2}$$

$$C_t \approx \frac{1}{\lambda\kappa^2}$$

where the last approximation follows by noting that $M_t$ is approximately the same as $M_t - \Delta t$.

Denote by $C_t^* = 1/(\lambda\kappa^2)$ this critical value of capital.

The evolution of capital follows a quadratic fractional transformation. In particular, its quadratic form is determined as the product of two affine ones. Cambini, Crouzeix, and Martein (2002) show that such a transformation is pseudoconvex in certain circumstances. The evolution of the capital can also be re-written as follows.

$$C_{t+\Delta t} = C_t + \frac{C_t\kappa\Delta t\left(R_{t;t+\Delta t} - \sigma_{t+\Delta t} + C_t\kappa\lambda\frac{(M_t-\Delta t)}{M_t}\right)}{M_t - C_t\kappa^2\lambda(M_t - \Delta t)}$$

$$= \frac{C_t\kappa\Delta t\left(R_{t;t+\Delta t} - \sigma_{t+\Delta t} + C_t\kappa\lambda\frac{(M_t-\Delta t)}{M_t}\right) + C_t\left(M_t - C_t\kappa^2\lambda(M_t - \Delta t)\right)}{M_t - C_t\kappa^2\lambda(M_t - \Delta t)}$$

$$= \frac{C_t\kappa\Delta t\left(R_{t;t+\Delta t} - \sigma_{t+\Delta t}\right) + C_t^2\kappa^2\lambda\frac{M_t-\Delta t}{M_t} + C_tM_t - C_t^2\kappa^2\lambda(M_t - \Delta t)}{M_t - C_t\kappa^2\lambda(M_t - \Delta t)}$$

$$= \frac{C_t\left(\kappa\Delta t\left(R_{t;t+\Delta t} - \sigma_{t+\Delta t}\right) + M_t\right) + C_t^2\kappa^2\lambda(M_t - \Delta t)\left(\frac{\Delta t}{M_t} - 1\right)}{M_t - C_t\kappa^2\lambda(M_t - \Delta t)}$$

Though this model is expressed in terms of derivatives and assumes a constant exposure, it is representative of a wider class of possible models that result in hedge funds buying small amounts of some securities for poor future arbitrage returns to achieve larger current mark-to-market returns.

## SIMULATIONS AND DISCUSSION

Let us consider a particular numerical example of this model. Suppose we are looking at an amalgamation of a group of hedge funds totaling $C_0$ = \$1 billion in capital, and that this capital is targeting κ = 10% of its capital as exposure to variance swaps. Hence, initially it has $100 of vega exposure.

(I use the term "vega notional" for convenience. The exact terminology would be "variance swap notional" and both the realized and implied marks would be in variance terms, not volatility, but the discussion flows easier if the terminology is that of volatility swaps. An alternative way of reading this is to assume that volatility swaps are decomposable.)

Suppose finally that the default maturity of new swaps is 5 years and that a $1 million purchase of new swaps increases all implieds by λ = 0.05 volatility points per $1 million. The particular numbers do not matter since similar results follow at some point so long as λ > 0 and κ > 0. Nevertheless, these numbers do represent ballpark estimates of reality. An online demonstration allows interaction and exploration of this model (Maymin, 2008b).

Figure 1 shows the evolution of the capital over time, where for convenience the capital, vega, and λ are all expressed in millions. This figure shows the prototypical evolution of capital under the model. Capital initially increases as the fund purchases more of the derivative both to replace its expiring exposure and to generate current mark-to-market profits at the expense of future arbitrage losses until the realized losses from expiring exposure gets so large that the fund is forced to sell and liquidate. The figure shows a gentle, smooth bubble.

Figure 1: Simulated Evolution of Capital Starting with Initial Capital of $1 billion

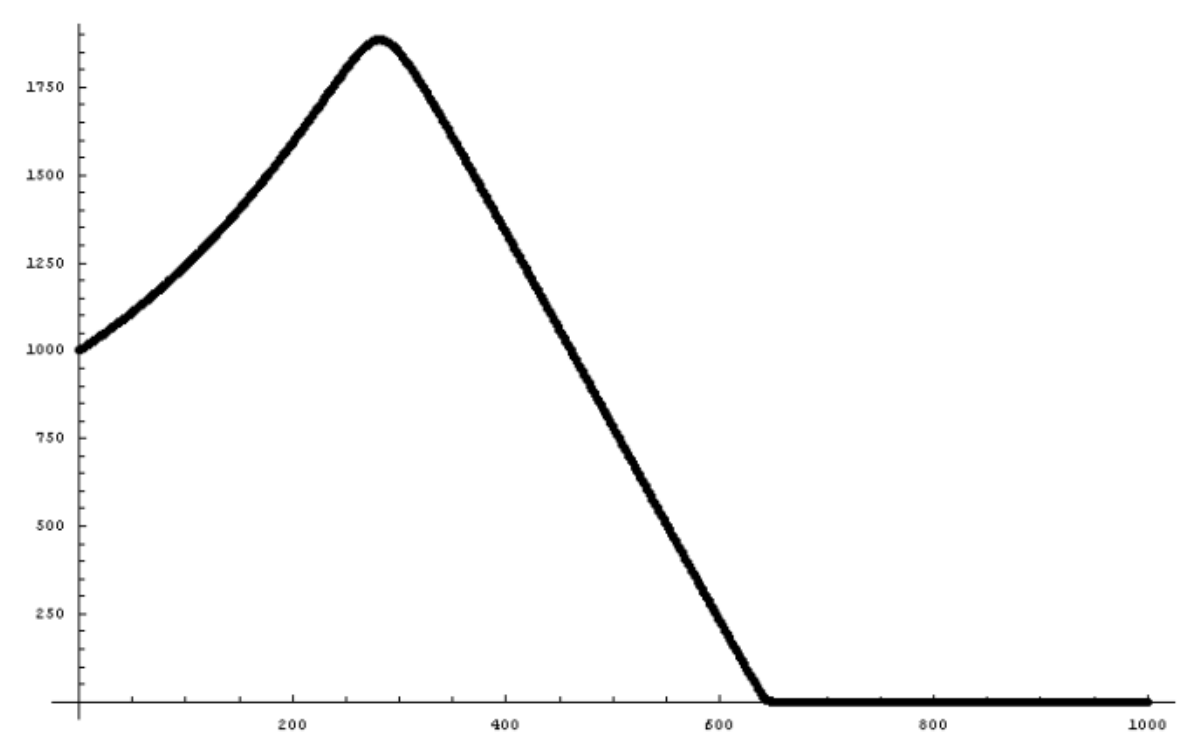

*This figure represents the evolution, in time steps of 1/100's of a year, of the capital of a hedge fund that invests 10% of its initial $1 billion in variance swaps that have a default maturity of 5 years and experience market impact equal to 0.05 variance points per $1 million notional trade. The initial implied is 20% and the realized is constant at 20%. The capital peaks above $1.75 billion after about three years (300 time steps), and then decreases rapidly until bankruptcy around year 6 (600 time steps).*

## <u>Discontinuities</u>

Recall the critical value of capital $C_t^* = \frac{1}{\lambda\kappa^2}$ at which discontinuities can occur and note that the critical value occurs when $C_t^* = \frac{1}{0.05/\$1{,}000{,}000\,(0.1)^2}$ = $2 billion. The capital never reaches the critical value and so the evolution remains smooth. But if the initial capital were just one percent higher and started out at $C_0$ = $1.01 billion, then its evolution would experience several discontinuities, including a big gap down after around two and a half years. See Figure 2.

Figure 2: Simulated Evolution of Capital Starting with Initial Capital of \$1.01 billion

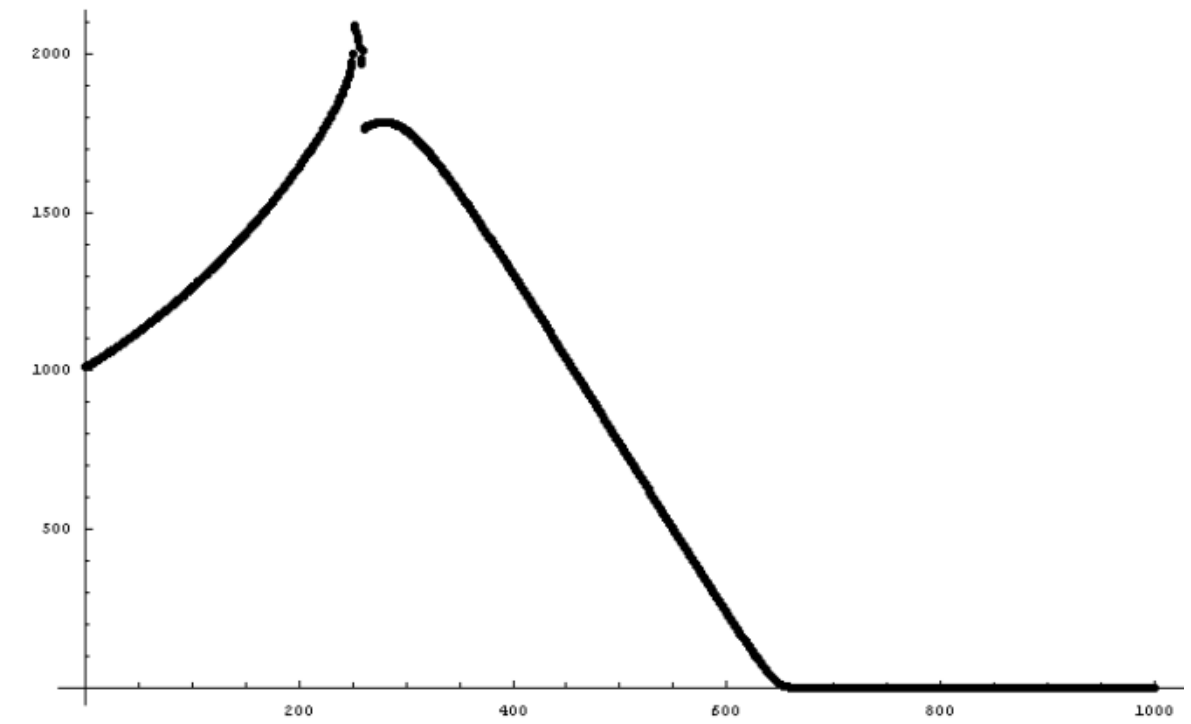

*This figure represents the evolution, in time steps of 1/100's of a year, of the capital of a hedge fund that invests 10% of its initial \$1.01 billion in variance swaps that have a default maturity of 5 years and experience market impact equal to 0.05 variance points per \$1 million notional trade. The initial implied is 20% and the realized is constant at 20%. The capital peaks above \$2 billion after about two and a half years (250 time steps), and then gaps down and decreases until bankruptcy around year 6 (600 time steps).*

Not all evolutions are necessarily bubbles either. If the initial capital were just two million dollars higher, a mere 0.20% higher than the previous example, so that $C_0$ = \$1.012 billion, then the capital of the hedge fund would continue to grow without bound. See Figure 3.

Figure 3: Simulated Evolution of Capital Starting with Initial Capital of \$1.012 billion

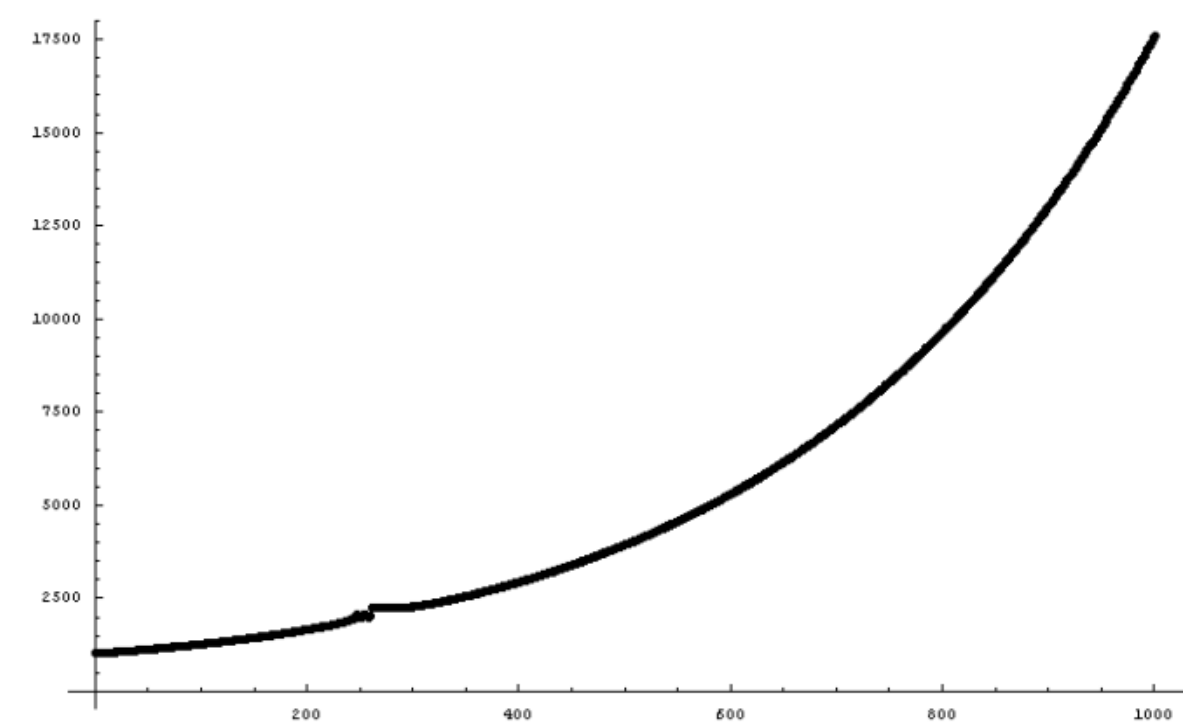

*This figure represents the evolution, in time steps of 1/100's of a year, of the capital of a hedge fund that invests 10% of its initial \$1.012 billion in variance swaps that have a default maturity of 5 years and experience market impact equal to 0.05 variance points per \$1 million notional trade. The initial implied is 20% and the realized is constant at 20%. The capital continues to increase indefinitely, reaching more than \$17.5 billion after ten years (1000 time steps).*

Additionally, it is not necessarily the case that there will only be one bubble. If we start from our base case and simply make the initial implied volatility 17% instead of 20%, so that the volatility appears to be 3% cheap to start, then the profits rise quickly, form a discontinuity, drop by approximately 75%, then start a new, slower, and more continuous bubble. See Figure 4.

Figure 4: Simulated Evolution of Capital Starting with Initial Implied Volatility of 17%

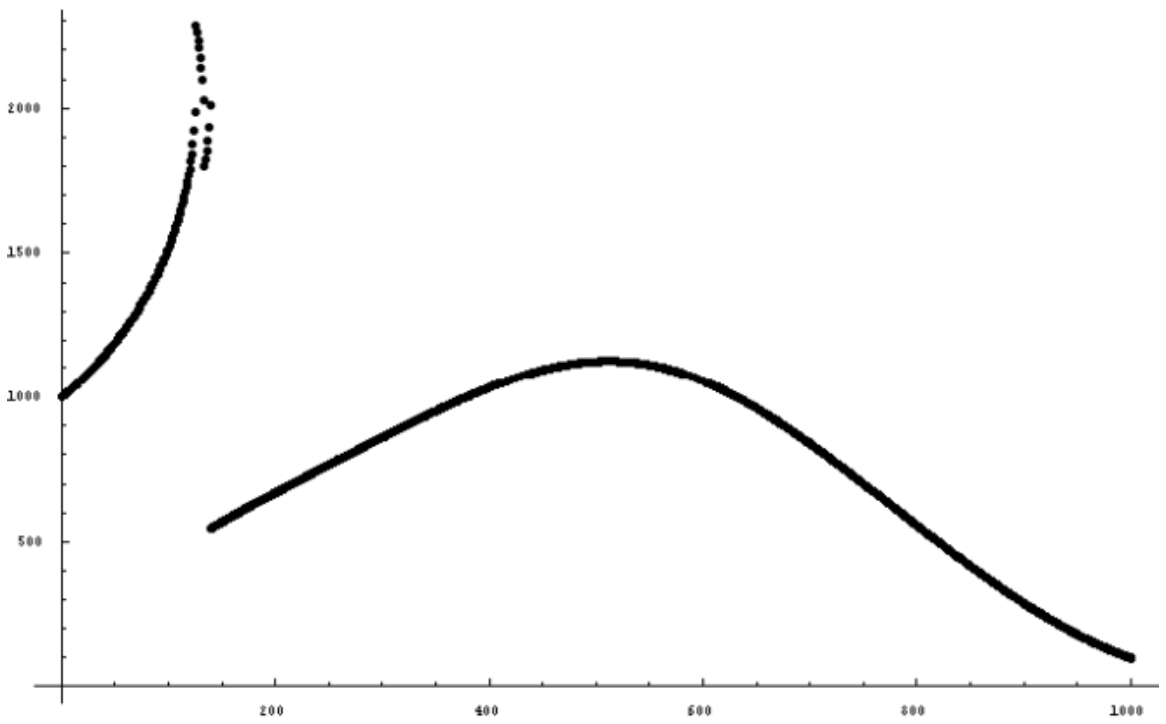

*This figure represents the evolution, in time steps of 1/100's of a year, of the capital of a hedge fund that invests 10% of its initial $1 billion in variance swaps that have a default maturity of 5 years and experience market impact equal to 0.05 variance points per $1 million notional trade. The initial implied is 17% and the realized is constant at 20%. The capital peaks at nearly $2.5 billion after about one and a half years (150 time steps), and then gaps down to about $500 million in capital. From there, it follows a smooth bubble back up to about $1 billion in year five (500 time steps) and decreases gradually towards bankruptcy.*

Because the evolution around the critical value is so fragile, different discontinuities can result from even slight changes to the time step or other parameters.

## Possible Extensions

The model can easily be extended to allow for alternative representations of market impact. For example, instead of a linear market impact model, we could use one whose impact is measured by the square root of the traded vega. However, that removes neither the jumps nor the possibility of infinite wealth, and apart from changing the exact times of bubbles and jumps, it does not aid in the intuition behind the model.

Similarly, the model can be extended to allow the market impact to dissipate over time, or to affect the realized portion as well as the implied portion. Indeed, there is good reason to assume that an increase in the implied will spill over into the realized simply because of the hedge fund's increased hedging activity. As more hedge funds buy volatility, volatility tends to dampen under their hedging. As more hedge funds sell volatility, volatility tends to increase under their hedging. Still, these enhancements to the model simply mitigate the effect over time but do not remove the fundamental discontinuities.

The model assumptions of direct decomposability and constant exposure may be relaxed. For example, mortgage-backed securities are not typically decomposable, but they do experience time decay in a non-linear way. Funds also may have a range of exposure they are willing to bear. The pure linearity of decomposition here is just a convenient approximation to the aging process of virtually all non-perpetual securities.

Similarly, though the illustrative example presented here was of a hedge fund trading volatility, the model applies equally well to a group or sector of hedge funds, or pension funds, or sovereign wealth funds, where allocations as a percentage of capital can be sticky even when the individual entities may not be because the risk of the entire securities held by those kinds of entities may be a fixed percentage, even though the individual entities allocate it different among themselves. As an example, by some anecdotal estimates, about 80 percent of the convertible bonds that were issued in the late 1990s and early 2000s were held by hedge funds; even if no particular hedge fund tried to hold a fixed proportion, the market was such that the overall amount held by hedge funds in total remained approximately constant.

## CONCLUSION

What are the potential long term effects of consistently propping up?

I present a model of financial entities seeking to maintain a constant exposure to decaying securities that engage in propping up and I find that they can generate a bubble. In most cases, the bubble can be a smooth runup followed by a smooth dropdown, but around certain critical capital values, discontinuities can result. Furthermore, the critical capital value depends only on two parameters: the market impact of trading new derivatives, and the proportional target exposure.

The implications for investors are that realized lower market volatility in such markets may be a temporary illusion hiding the possibility of a chaotic crash, and that the amount of capital committed to the strategies can provide valuable information to the extent the critical values can be calculated.

With the current environment of global government intervention into private markets, whether by purchasing outright equity stakes or establishing a portfolio of bad assets, the long-term consequences may not be adequately addressed or even considered. This is, after all, an "emergency" situation. But the proposed cure may be worse than the disease, and if the level of governmental intervention reaches the critical value, defined above as a function of the market impact and the allocation percentage, it is essentially unpredictable what may result.

## ACKNOWLEDGMENTS AND CONTACT INFORMATION

*I am grateful to Andrew Derry and an anonymous referee at the Wolfram Demonstrations Project for their comments.* *My email address is pmaymin@poly.edu.*

## BIOGRAPHY

Dr. Philip Maymin is an assistant professor of finance and risk engineering at the Polytechnic Institute of New York University. He holds a Ph.D. in finance from the University of Chicago, an M.S. in applied mathematics from Harvard University, and a B.A. in computer science from Harvard University. He is a Policy Scholar for the Yankee Institute for Public Policy, an award-winning journalist, and the author of *Yankee Wake Up*. He has also been a portfolio manager at Long-Term Capital Management, Ellington Management Group, and his own hedge fund, Maymin Capital Management. He was the 2006 Libertarian candidate for U.S. House of Representatives in Connecticut's fourth Congressional district.